# Broken rotational symmetry in the pseudogap phase of a high-$T_c$ superconductor

R. Daou [1,†], J. Chang [1], David LeBoeuf [1], Olivier Cyr-Choinière [1], Francis Laliberté [1], Nicolas Doiron-Leyraud [1], B. J. Ramshaw [2], Ruixing Liang [2,3], D. A. Bonn [2,3], W. N. Hardy [2,3] & Louis Taillefer [1,3]

*1 Département de physique and RQMP, Université de Sherbrooke, Sherbrooke, Québec J1K 2R1, Canada*

*2 Department of Physics and Astronomy, University of British Columbia, Vancouver, British Columbia V6T 1Z1, Canada*

*3 Canadian Institute for Advanced Research, Toronto, Ontario M5G 1Z8, Canada*

**The nature of the pseudogap phase is a central problem in the quest to understand high-$T_c$ cuprate superconductors[1]. A fundamental question is what symmetries are broken when that phase sets in below a temperature $T^*$. There is evidence from both polarized neutron diffraction[2,3] and polar Kerr effect[4] measurements that time-reversal symmetry is broken, but at temperatures that differ significantly. Broken rotational symmetry was detected by both resistivity[5] and inelastic neutron scattering[6,7,8] at low doping and by scanning tunnelling spectroscopy[9,10] at low temperature, but with no clear connection to $T^*$. Here we report the observation of a large in-plane anisotropy of the Nernst effect in $YBa_2Cu_3O_y$ that sets in precisely at $T^*$, throughout the doping phase diagram. We show that the CuO chains of the orthorhombic lattice are not responsible for this anisotropy, which is therefore an intrinsic property of the $CuO_2$ planes. We conclude that the pseudogap phase is an electronic state which strongly breaks four-fold rotational symmetry. This narrows the range of possible states considerably, pointing to stripe or nematic orders[11,12].**

† Present address : Max Planck Institute for Chemical Physics of Solids, 01187 Dresden, Germany

We have measured the Nernst coefficient $\nu(T)$ of the high-$T_c$ superconductor $YBa_2Cu_3O_y$ (YBCO) as a function of temperature up to $\sim 300$ K for a hole concentration[13] (doping) ranging from $p = 0.08$ to $p = 0.18$, in untwinned crystals where the temperature gradient $\Delta T$ was applied along either the $a$-axis or the $b$-axis of the orthorhombic plane. In Fig. 1, a typical data set is seen to consist of two contributions: 1) a positive, strongly field-dependent contribution due to superconducting fluctuations[14,15,16]; 2) a field-independent contribution due to normal-state quasiparticles[17], which drops from small and positive to large and negative with decreasing temperature. We define as $T_\nu$ the temperature below which $\nu / T$ starts its downward drop. In Fig. 2, we plot $T_\nu$ as a function of doping. We also plot $T_\rho$, the temperature below which the in-plane resistivity $\rho(T)$ of YBCO deviates downward from its linear temperature dependence at high temperature, a standard definition of the pseudogap temperature $T^*$ (refs. 18, 19). We see that $T_\nu = T_\rho$, within error bars, as also found in a recent study on YBCO films[20]. We also see that $T_\nu$ obtained with $\Delta T \parallel a$ is the same as $T_\nu$ obtained with $\Delta T \parallel b$, within error bars. We therefore conclude that the drop in the quasiparticle Nernst signal to large negative values is a signature of the pseudogap phase, detectable up to the highest measured doping, $p = 0.18$.

In Fig. 3, we see that the dip in $\nu / T$ between $T_c$ and $T_\nu$ gets deeper with decreasing $p$ as the separation between $T_c$ and $T_\nu$ grows (Fig. 2). This characteristic dip is hugely anisotropic, being roughly 10 times deeper when $\Delta T \parallel b$. In Fig. S6, the Nernst anisotropy is plotted as a ratio, seen to reach $\nu_b / \nu_a \approx 7$ at 90 K for $p = 0.12$. To our knowledge, this is the largest in-plane anisotropy reported in any macroscopic physical property of any high-$T_c$ superconductor[12]. In Fig. 4a, a plot of the anisotropy difference $D(T) \equiv (\nu_a - \nu_b) / T$ reveals that the onset of this $a$-$b$ anisotropy coincides with $T_\nu$, showing that it is a property of the pseudogap phase, since $T_\nu = T^*$. In Fig. 4b, we plot the difference normalized by the sum $S(T) \equiv -(\nu_a + \nu_b) / T$; this relative anisotropy, $D(T) / S(T) = (v_b - v_a) / (v_b + v_a)$, can be viewed as a Nernst-derived nematic order parameter, in analogy with that defined from the resistivity[21].

In the orthorhombic crystal structure of YBCO, there are CuO chains along the $b$-axis, between the $CuO_2$ planes common to all cuprates. These one-dimensional chains can conduct charge, causing an anisotropy in the conductivity $\sigma$ such that $\sigma_b / \sigma_a > 1$. In principle these chains could also cause an anisotropy in $v$, but we next show that the chains make a negligible contribution to $v$. We first consider the low doping regime at $p = 0.08$ ($y = 6.45$), for which the anisotropy ratio of both $\sigma$ and $v$ is displayed in Fig. S6a. As established previously[5], the conductivity of chains decreases with decreasing $p$ until it becomes negligible by $p \approx 0.08$, as shown by the fact that $\sigma_b / \sigma_a \approx 1$ at high temperature. In that context of negligible chain conduction, a small rise in the anisotropy ratio $\sigma_b / \sigma_a$ with decreasing temperature is seen (Fig. S6a), convincing evidence of a state that breaks rotational symmetry, as previously reported[5]. The similar (but larger) rise in $v_b / v_a$ (Fig. S6a) is equivalent evidence of the same symmetry breaking. By contrast, at higher doping, such as $p = 0.12$ (Fig. S6b), $\sigma_b / \sigma_a$ now decreases upon cooling, the signature of chain-dominated conductivity[5], but the Nernst anisotropy still exhibits the same characteristic rise upon cooling as for $p = 0.08$. This shows that while chains now dominate the $\sigma$ anisotropy, they appear to have little impact on the $v$ anisotropy.

This is confirmed by a second test, where we greatly enhance the conductivity of chains while keeping the doping approximately constant. This is done by comparing samples with $y = 6.97$ ($p = 0.177$) to samples with $y = 6.998$ ($p = 0.180$). Because the density of oxygen vacancies in the chains is 3% vs 0.2%, respectively, the chain conductivity of the 6.998 samples is much larger, by a factor of 4 (see Fig. S8). The effect of this enhanced chain conductivity on the Nernst signal can be seen in the anisotropy difference $D(T)$, plotted in Fig. S9b. At $T > T_v$, it produces a temperature-dependent background in $D(T)$, falling with decreasing temperature, visible only in the 6.998 samples. At $T < T_\nu$, $D(T)$ increases in similar fashion for all samples (Fig. 4a): the effect of the pseudogap is clearly to increase $D(T)$. Now if the chains were responsible for this increase, we would expect the increase to be largest in the samples with the most highly conducting chains, namely the 6.998 samples. The opposite is true: below $T_\nu$, $D(T)$ is smallest for those samples (Fig. S9b). We conclude that chain conduction is not the cause of the pseudogap-related anisotropy in the Nernst coefficient.

This implies that the pseudogap phase breaks the four-fold rotational symmetry of the $CuO_2$ planes. Of course, the orthorhombic distortion of the $CuO_2$ planes caused by the CuO chains already breaks four-fold symmetry, and this "weak" symmetry breaking is necessary for any breaking of four-fold rotational symmetry to be observable macroscopically. In its absence, any spontaneous order would form domains and the associated in-plane anisotropy would be averaged out to zero over the volume of the sample[12]. The orthorhombic distortion plays the same role as an in-plane magnetic field in a ferromagnet or a metal with nematic order[21].

Broken rotational symmetry places a major constraint on the possible states that can be identified with the pseudogap phase. It favours "stripe-like" order – unidirectional modulations of the spin and / or charge density – or nematic order[11,12]. Recent calculations applied to cuprates confirm that stripe order can cause a major enhancement of the quasiparticle Nernst signal, with a sign that depends on the particular $\boldsymbol{Q}$ vector[22], and nematic order can produce a much larger anisotropy in $\nu$ than in $\sigma$ (ref. 23).

In $La_{2-x}Sr_xCuO_4$ (LSCO) doped with Nd or Eu, the quasiparticle Nernst signal also undergoes an enhancement (and sign change) below a temperature $T_\nu$ which is equal to $T_\rho$ (ref. 24), both temperatures decreasing monotonically with doping in a way which is very similar to YBCO (Fig. 2), with $T_\nu$ (and $T_\rho$) going to zero at a critical doping $p* \approx 0.24$ (ref. 25). In these materials, static long-range stripe order has been observed below a temperature $T_{CO}$ which decreases with doping and vanishes at $p \approx 0.25$ (ref. 26), with $T_{CO} \approx T_\nu / 2$ (ref. 27). The onset of stripe order causes a reconstruction of the Fermi surface at $T_{CO}$ (ref. 27) which, in Eu-doped LSCO at $p = 0.125$, shows up as a drop in the Hall coefficient $R_H(T)$ and the Seebeck coefficient $S(T)$ to negative values[27,28], starting at $T_{CO} \approx 80$ K. In YBCO at $p = 0.12$, the same drop is observed in both $R_H(T)$ and $S(T)$, at the same temperature[27,28]. The fact that $R_H$ and $S/T$ become deeply negative in the normal state at low temperature is ascribed to the formation of an electron pocket in the Fermi surface of YBCO (refs. 28, 29). All this argues strongly for stripe-like order in YBCO at low temperature (in the absence of superconductivity).

It therefore appears that the transformation of the electronic state in underdoped cuprates upon cooling proceeds in two stages: a first transformation at $T*$, where rotational symmetry is broken, and a second transformation at $T \approx T*/2$, where translational symmetry is broken (at least in the absence of superconductivity). The first regime may simply be a short-range / fluctuating precursor of the state at low temperature. This two-stage evolution is consistent with neutron scattering studies[8] of YBCO at low doping ($T_c = 35$ K; $p \approx 0.07$). Upon cooling from high temperature, an anisotropy in the spin fluctuation spectrum appears below 150 K or so, in the form of an incommensurability which grows with decreasing temperature, observed along the $a*$ axis but not the $b*$ axis[8]. At low temperature, static incommensurate spin-density-wave (SDW) order is seen[8]. Even though these particular observations are at a doping below the range of our investigations, the two-stage ordering sequence they reveal at $p \approx 0.07$ – anisotropic spin fluctuations followed by SDW order – is consistent with the two-stage process of symmetry breaking revealed by transport measurements at $p = 0.12$. (Note that the in-plane anisotropy of the spin fluctuation spectrum is present at least up to $p \approx 0.11$ (refs. 6, 7).) This type of ordering sequence – fluctuating to static stripe order – was proposed theoretically long ago[30].

**Acknowledgements** We thank K. Behnia, R.L. Greene, C. Kallin, S.A. Kivelson, A.J. Millis, C. Proust, S. Sachdev, A.-M.S. Tremblay and M. Vojta for stimulating discussions, and J. Corbin for his assistance with the experiments. JC was supported by fellowships from the Swiss National Science Foundation and the FQRNT. LT acknowledges support from the Canadian Institute for Advanced Research and funding from NSERC, FQRNT, CFI and a Canada Research Chair.

**Author Contributions** R. D., J. C., D. L., O. C-C., F. L. and N. D.-L. performed the Nernst and resistivity measurements; R. D., J. C. and D. L. analyzed the Nernst data; B. J. R., R. L., D. A. B. and W. N. H. prepared the samples at UBC (crystal growth, annealing, de-twinning, contacts); L. T. supervised the project and wrote the manuscript.

**Author Information** Correspondence and requests for materials should be addressed to L.T. (louis.taillefer@physique.usherbrooke.ca).

## Figure 1 | Nernst coefficient.

**a,** Nernst coefficient $v$ of YBCO, plotted as $v / T$ vs $T$, for different values of the magnetic field $B$ as indicated. The sample is a single crystal with oxygen content $y = 6.67$ and a superconducting transition temperature $T_c(B=0) = 66.0$ K (vertical line), corresponding to a hole concentration (doping) of $p = 0.12$ (ref. 13). The thermal gradient is along the $b$-axis of the orthorhombic structure, the transverse Nernst voltage is measured along the $a$-axis, with the field applied along the $c$-axis. $v(T)$ is seen to consist of two contributions: 1) a field-independent contribution, attributed to quasiparticles, which is small and positive at high temperature and becomes large and negative upon cooling; 2) a positive, strongly field-dependent contribution, attributed to superconducting fluctuations, which causes $v$ to rise sharply as the temperature gets close to $T_c$. **b,** Zoom at high temperature, showing where $v / T$ starts to fall below its flat, small and positive value, at the onset temperature $T_v$ (arrow). The data for all samples are shown in the Supplementary Information (Figs. S1, S2 and S3) and the values of $T_v$ are listed in Table S1. Our data are consistent with published YBCO data[14,15,16] wherever they overlap (in temperature and doping). In particular, a negative quasiparticle contribution has been observed for $p \approx 0.1$ - $0.15$, in measurements mostly done on twinned crystals[14,15,16].

**Figure 2 | Phase diagram.**

Temperature – doping phase diagram of YBCO showing the superconducting phase (SC) below the transition temperature $T_c$ (diamonds; from ref. 13). Two characteristic temperatures are plotted: 1) $T_v$ (circles), the onset of the drop in the quasiparticle Nernst signal, as defined in Fig. 1b (and Figs. S1 and S2), for $\Delta T \parallel a$ (blue) and $\Delta T \parallel b$ (red); 2) $T_\rho$ (green squares), the onset of the drop in the resistivity $\rho(T)$ from its linear temperature dependence at high temperature, as defined in Fig. S4 (using data from ref. 19). This is the standard definition of the pseudogap onset temperature $T^*$ in YBCO (ref. 18). Error bars on $T_v$ and $T_\rho$ indicate the uncertainty in locating the temperature below which $v / T$ and $\rho(T)$ start to drop from their respective behaviour at high temperature (see Supplementary Information). The dashed line is a guide to the eye. Within error bars, we find that $T_v = T_\rho$, showing that the drop in $v / T$ signals the onset of the pseudogap phase. Note that in $v_b / T$ this signature remains clearly visible up to the highest measured doping ($p = 0.18$) (Fig. S1f), while it is no longer detectable in $v_a / T$ or $\rho$ beyond $p \approx 0.15$ (Fig. S2).

**Figure 3 | Comparison of $v_a$ and $v_b$ .**

Temperature dependence of $v / T$ normalized to $T_v$ at various dopings as indicated, for: **a,** $\Delta T \parallel a$; **b,** $\Delta T \parallel b$. The $T_v$ values are those listed in Table S1 of the Supplementary Information, and plotted in Fig. 2. Note that the vertical range is 10 times larger in **b**, showing that the negative quasiparticle Nernst signal is an order of magnitude larger for $\Delta T \parallel b$.

**Figure 4 | Anisotropy of the Nernst signal.**

Difference in the Nernst signal between $\Delta T \parallel a$ and $\Delta T \parallel b$, defined as $D(T) \equiv (v_a - v_b) / T$. **a,** plotted as $D(T) - D(T_v)$ vs $T / T_v$ , for dopings as indicated (see full un-normalized data in Fig. S5). $T_v$ is that obtained for the $b$-axis samples (Fig. S1 and Table S1). The onset of the pseudogap phase at $T_v$ is seen to cause a fairly uniform rise in the anisotropy. **b,** plotted as $[D(T) - D(T_v)] / [S(T) - S(T_v)]$ vs $T$ for $p = 0.12$ (open black circles), where $S(T) \equiv - (v_a + v_b) / T$. This ratio becomes equal to $(v_b - v_a) / (v_b + v_a)$ (full red circles) at low temperature, when $D(T) >> |D(T_v)|$ and $S(T) >> |S(T_v)|$. The latter ratio can be viewed as a nematic order parameter (see ref. 21). (Note that because $(v_b - v_a)$ changes sign near 150 K, it is meaningless to plot $(v_b - v_a) / (v_b + v_a)$ beyond 120 K or so; see Fig. S7). The error bar on the absolute value of $(v_b - v_a) / (v_b + v_a)$ (shown at 90 K) comes from the separate uncertainties of $\pm$ 10 % on $v_b$ and $v_a$ (see Supplementary Information). The dotted line shows a simple parabolic dependence.

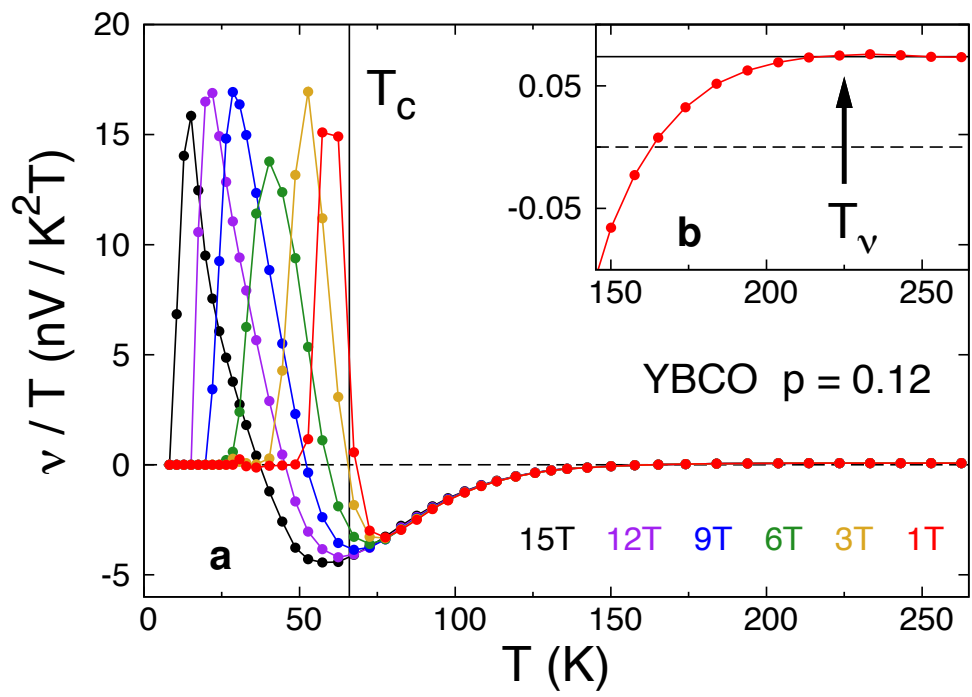

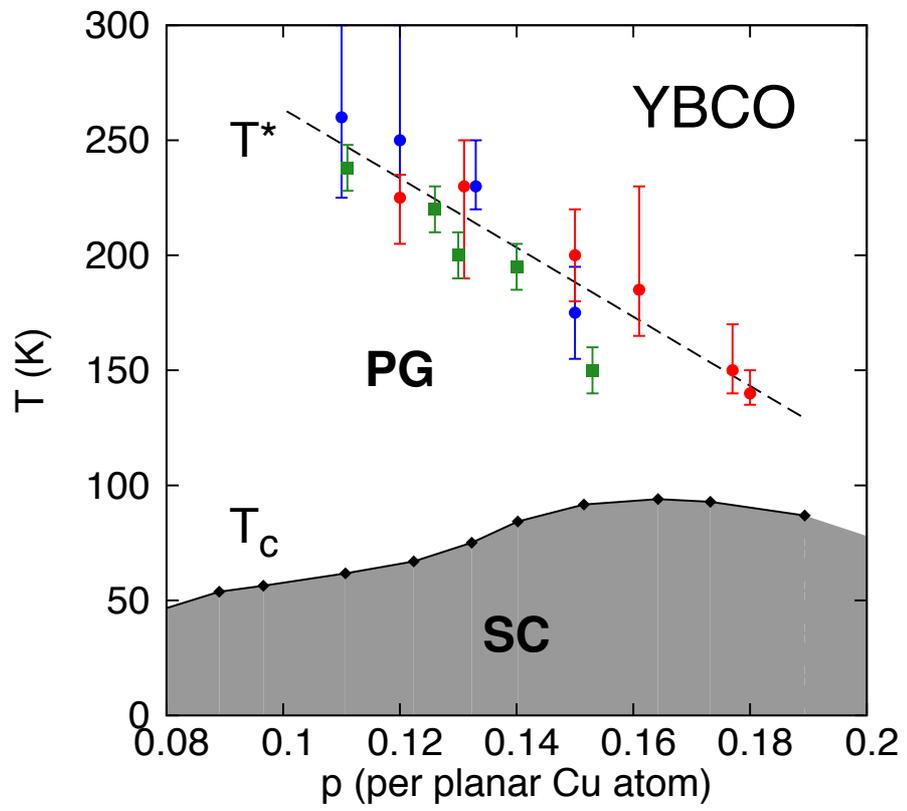

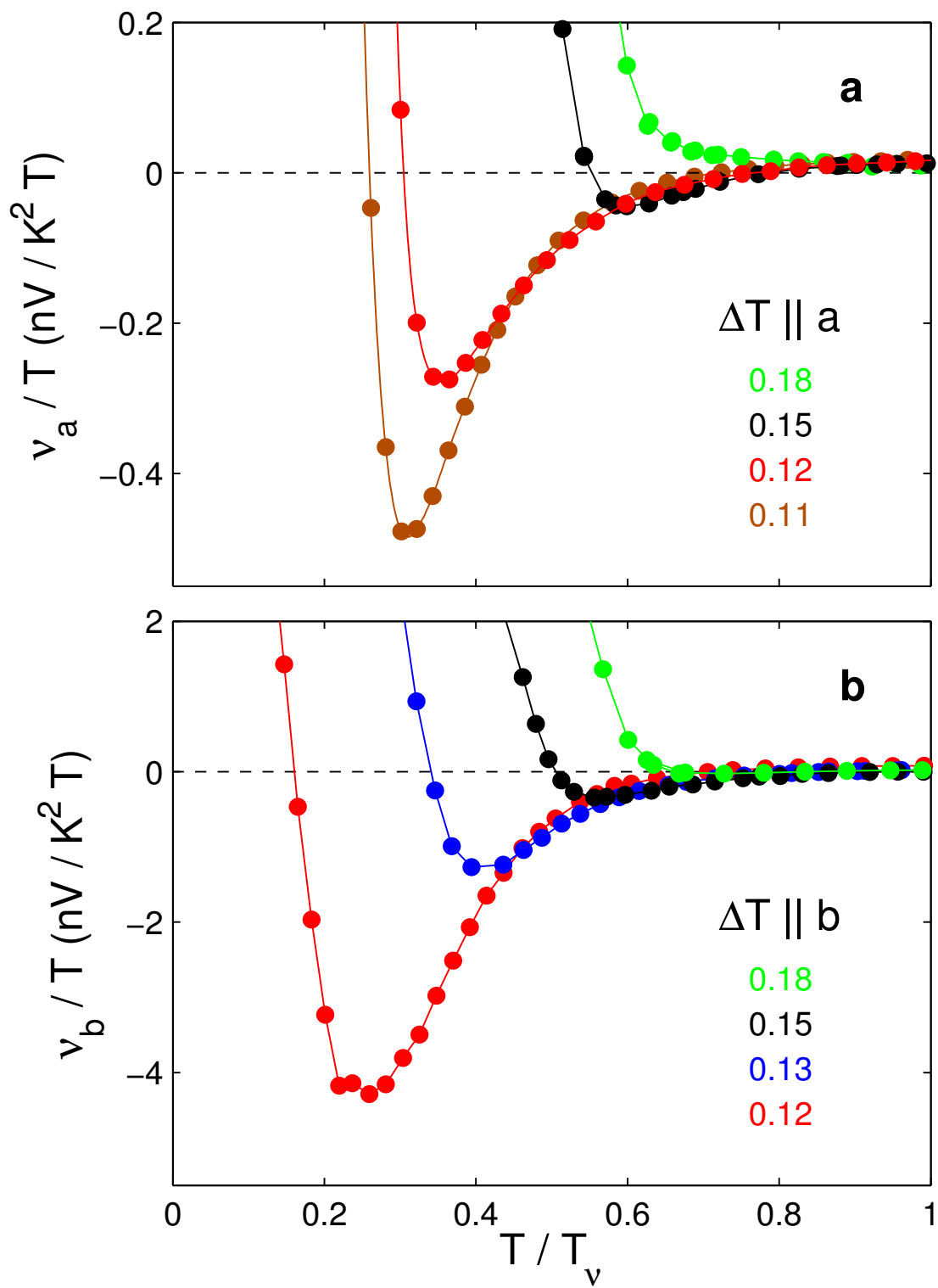

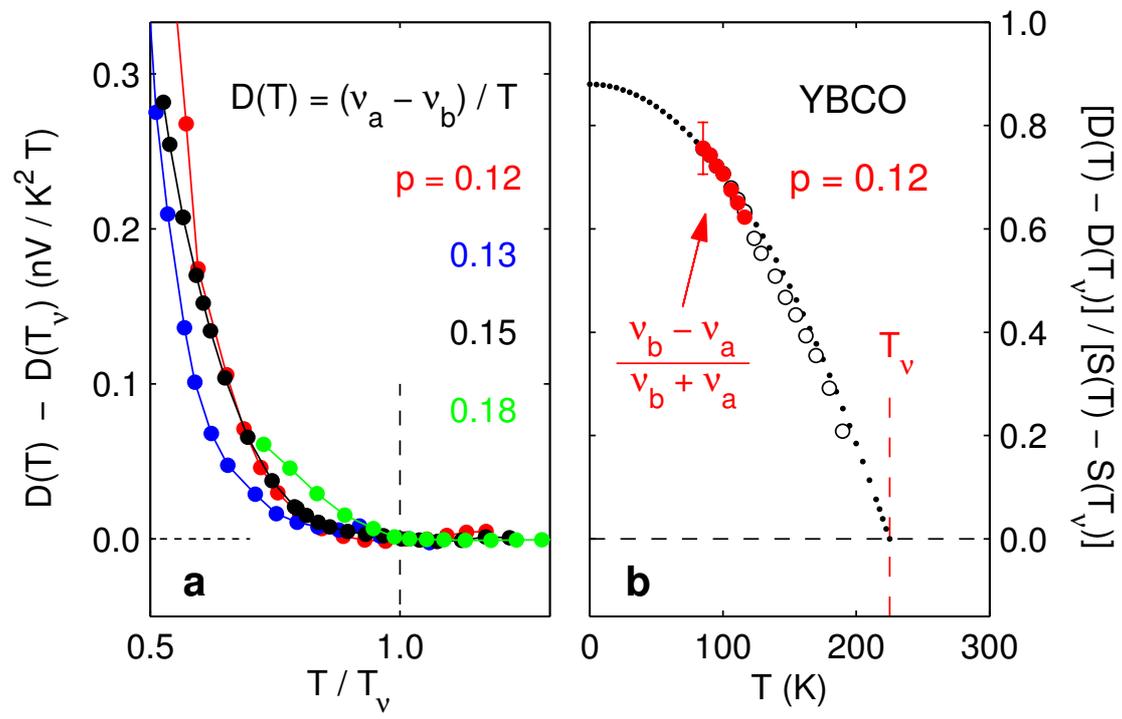

# THE NERNST EFFECT

The Nernst effect is the development of a transverse electric field $E_y$ across the width ($y$-axis) of a metallic sample when a temperature gradient $\partial T / \partial x$ is applied along its length ($x$-axis) in the presence of a transverse magnetic field $B$ (along the $z$-axis). Two mechanisms can give rise to a Nernst signal[17], $N \equiv E_y / ( \partial T / \partial x )$: superconducting fluctuations[14], which give a positive signal, and charge carriers (quasiparticles), which can give a signal of either sign. At low temperature, the magnitude of the quasiparticle Nernst signal is given approximately by[17]:

$$\nu / T \approx \pi^2 / 3 \, ( \, k_B{}^2 / e \, ) \, ( \, \mu / \varepsilon_F \, ) \quad , \qquad (1)$$

where $\nu \equiv N / B$ is the Nernst coefficient, $T$ is the temperature, $k_B$ is Boltzmann's constant, $e$ is the electron charge, $\mu$ is the carrier mobility and $\varepsilon_F$ the Fermi energy. Eq. (1) works remarkably well as a universal expression for the Nernst coefficient of metals at $T \to 0$, accurate within a factor two or so in a wide range of materials[17]. It explains why a phase transition which reconstructs a large Fermi surface into small pockets (with small $\varepsilon_F$) can cause a major enhancement of $\nu$. The heavy-fermion metal $URu_2Si_2$ provides a good example of this phenomenon. As the temperature drops below its transition to a semi-metallic state at 17 K, the carrier density $n$ (or $\varepsilon_F$) falls and the mobility rises, both by roughly a factor 10, and $\nu / T$ increases by a factor 100 or so[31]. Note that the electrical resistivity $\rho(T)$ is affected only weakly by these dramatic changes[32], since mobility and carrier density are modified in ways which compensate in the conductivity $\sigma = 1/ \rho = n \, e \, \mu$. This is why the Nernst effect is a vastly more sensitive probe of electronic transformations, such as density-wave transitions, than the resistivity. Here we use it to probe the pseudogap phase of a high-$T_c$ superconductor.

# EXPERIMENTAL DETAILS

**Crystal structure**. The hole-doped cuprate YBa$_2$Cu$_3$O$_y$ (YBCO) has a lattice structure of orthorhombic symmetry, made of CuO$_2$ planes stacked in pairs (bi-layers) along the $c$-axis, with non-equivalent $a$ and $b$ lattice parameters in the orthorhombic plane. In the middle of the separation between adjacent CuO$_2$ bi-layers, there is a layer of one-dimensional CuO chains running along the $b$-axis. The oxygen content of these chains can be varied by annealing, from full at $y = 7.0$ to empty at $y = 6.0$. For $y > 6.5$ or so, the chains conduct, at least at high temperature, causing an anisotropy in the DC conductivity $\sigma$, typically in the range $\sigma_b / \sigma_a = 1 - 2.5$ (ref. 5).

**Samples**. Our YBCO samples are fully detwinned crystals grown in non-reactive BaZrO$_3$ crucibles from high-purity starting materials (see ref. 33). The samples are uncut, unpolished thin platelets, whose transport properties are measured via gold evaporated contacts (of resistance $< 1 \ \Omega$), in a six-contact geometry. Typical sample dimensions are 20-50 $\times$ 500-800 $\times$ 500-1000 $\mu$m$^3$ (thickness $\times$ width $\times$ length).

**Estimates of hole concentration**. The hole concentration (doping) $p$ in YBCO was determined from a relationship between $T_c$ and the $c$-axis lattice constant[13]. The value of $T_c$ for each sample was defined as the temperature where its resistance goes to zero. The $T_c$ values and corresponding $p$ values are listed in Table S1 for the 14 samples studied here.

| y | ΔT | $T_c$ (K) | p | $T_v$ (K) |
|---|---|---|---|---|
| 6.45 | a | 45 | 7.8 | ------- |
| 6.45 | b | 45 | 7.8 | ------- |
| 6.54 | a | 61.5 | 11.0 | 260 |
| 6.67 | a | 66 | 12.0 | 250 |
| 6.67 | b | 66 | 12.0 | 225 |
| 6.75 | a | 75 | 13.2 | 230 |
| 6.75 | b | 75 | 13.2 | 230 |
| 6.86 | a | 91 | 15.0 | 175 |
| 6.86 | b | 91 | 15.0 | 200 |
| 6.92 | b | 93.5 | 16.1 | 185 |
| 6.97 | a | 91.5 | 17.7 | --------- |
| 6.97 | b | 91.5 | 17.7 | 150 |
| 6.998 | a | 90.5 | 18.0 | --------- |
| 6.998 | b | 90.5 | 18.0 | 140 |

**Table S1 | Sample characteristics.**

Oxygen content *y*, temperature gradient direction, $T_c$, doping *p* and $T_v$ for each of the 14 YBCO samples measured in this study. See text for definitions of $T_c$, *p* and $T_v$. The error bar on $T_c$ is typically ± 0.2 K. The error bar on $T_v$ is shown in Figs. S1 and S2.

**Measurement of the Nernst coefficient**. The Nernst signal was measured by applying a steady heat current through the sample (along the *x*-axis). The longitudinal thermal gradient was measured using two uncalibrated Cernox chip thermometers (Lakeshore), referenced to a further calibrated Cernox. The temperature of the experiment was stabilized at each point to within ± 10 mK. The temperature and voltage were measured with and without applied thermal gradient (Δ*T*) for calibration. The magnetic field *B*, applied along the *c*-axis (*B* ∥ *z*), was then swept, with the heat on, from − 15 to + 15 T at 0.4 T / min, continuously taking data. The thermal gradient was monitored continuously and remained constant during the course of a sweep. The Nernst coefficient (*N*) was extracted from that part of the measured voltage which is anti-symmetric with respect to the magnetic field:

$$N = E_y / ( \partial T / \partial x ) = [ \Delta V_y(B) / \Delta T_x - \Delta V_y(\text{-}B) / \Delta T_x ] ( L / 2\,w ) \quad ,$$

where Δ*V* is the difference in the voltage measured with and without thermal gradient. *L* is the length (between contacts along the *x*-axis) and *w* the width (along the *y*-axis) of the sample. This anti-symmetrization procedure removes any longitudinal thermoelectric contribution from the sample and a constant background from the measurement circuit. The uncertainty on *N* comes mostly from the uncertainty in measuring *L* and *w*, giving a typical error bar of ± 10 % on ν.

The Nernst effect was measured in 14 YBCO samples. The raw data are shown in Figs. S1, S2 and S3. All the Nernst data displayed here (whether in the main article or in this Supplementary Information) are for an applied magnetic field *B* = 15 T, except for the *p* = 0.13 samples (both *a*-axis and *b*-axis), where *B* = 10 T. Note that the quasiparticle Nernst coefficient of interest here is completely independent of magnetic field. For only one curve, the *p* = 0.12 *a*-axis curve in Fig. 3a, we used data taken at a different field, namely *B* = 3 T. The reason is cosmetic: to make the rise due to the superconducting contribution in the *p* = 0.12 data well separated from the rise in the *p* = 0.11 data.

## DATA ANALYSIS

**Definition of the pseudogap temperature $T^*$.** Following the standard definition[18,19], we define the pseudogap temperature in YBCO to be the temperature $T_\rho$ below which the $a$-axis resistivity drops below its linear temperature dependence at high temperature. In Fig. S4, an example is given for $p = 0.13$, both from our own data and from published data[19]. In Fig. 2, we plot $T_\rho$ for different dopings (using data from ref. 19).

**Definition of $T_\nu$ .** We define $T_\nu$ as the temperature below which $\nu / T$ falls below its maximal value at high temperature, as shown in Figs. S1 and S2. Because this is not a sharp transition but a smooth crossover, estimates of $T_\nu$ have some uncertainty, dependent also on the noise level of the data. In Figs. S1 and S2, we show what we feel are reasonable uncertainties on $T_\nu$ for each sample. These are then plotted in Fig. 2. In Fig. S4d, we show how resistivity and Nernst coefficient both deviate simultaneously from their linear high-temperature behaviour. In Fig. S4d and Fig. 2, we see that $T_\nu$ and $T_\rho$ are equal within error bars, showing that the drop in $\nu / T$ is caused by the onset of the pseudogap phase. We also show that within error bars $T_\nu$ is the same for $\Delta T \parallel a$ and $\Delta T \parallel b$. With increasing $p$, as $T_\nu$ and $T_c$ come together, the dip in $\nu / T$ becomes shallower (Fig. 3). For $\Delta T \parallel a$, it can no longer be resolved at $p = 0.177$ (Fig. S2e). However, because it is much more pronounced for $\Delta T \parallel b$, roughly by a factor 10 (Fig. S3), the dip remains clearly visible in all $b$-axis samples, up to and including $p = 0.18$ (Fig. S1).

**Anisotropy of the Nernst signal.** The anisotropy is obtained directly from the raw Nernst signals $\nu_a$ ($\Delta T \parallel a$) and $\nu_b$ ($\Delta T \parallel b$) measured on a pair of de-twinned crystals prepared together, in identical fashion and hence with the very same doping (Fig. S3). It is plotted as a difference $D(T) \equiv (\nu_a - \nu_b) / T$ in Figs. 4a and S5, as a ratio $\nu_b / \nu_a$ in Fig. S6, and as a fraction $(\nu_b - \nu_a) / (\nu_b + \nu_a)$ in Figs. 4b and S7.

# THE ROLE OF CuO CHAINS

Here we summarize the four arguments put forward to rule out chain conductivity as the cause of the large anisotropy in the Nernst signal below $T_v$.

The first argument is that chain-related anisotropy, as manifest in the conductivity, *decreases* with decreasing temperature below 150 K, at all dopings (see ref. 5). By contrast, the Nernst anisotropy *grows* with decreasing temperature, at all dopings.

The second argument is that the Nernst anisotropy undergoes a pronounced increase starting at $T_v$, being very small and temperature-independent above $T_v$ (Fig. S5). By contrast, chain conductivity is either entirely unaffected by the onset of the pseudogap phase (as in the $y = 6.998$ samples; see Fig. S9a) or possibly suppressed (see ref. 5).

The third argument is that the Nernst anisotropy remains large even when chain conductivity has been essentially switched off, as in the $p = 0.08$ samples where $\sigma_b / \sigma_a$ has become negligibly small even at room temperature (see Fig. S6a and ref. 5).

The fourth and most compelling argument is that the Nernst anisotropy is not enhanced by making the conductivity of chains 4 times larger at a nearly identical doping, as in the $y = 6.998$ samples vs the $y = 6.97$ samples (see Figs. S8 and S9). In fact, the reverse is true: the very high chain conductivity in 6.998 causes an anisotropy in the Nernst signal which is opposite to the pseudogap-related anisotropy seen in all samples. Indeed, the total Nernst signal is made *less* anisotropic below $T_v$, not more, by making the chains more conducting, *e.g.* $v_b / v_a \approx 1$ at 100 K (see Fig. S3f). As a result of this compensating effect of chains, the anisotropy difference in the 6.998 samples is smaller below $T_v$ than it would otherwise be (see Fig. S9). Correcting for this chain-related background yields a universal rate of growth in the anisotropy below $T_v$ (Figs. 4a and S9c).

## Figure S1 | Nernst coefficient of *b*-axis samples (Δ*T* // *b*).

**a – f,** Nernst coefficient $v$ of *b*-axis YBCO samples (Δ*T* // *b*) measured in a magnetic field $B = 15$ T (10 T for the $p = 0.13$ sample in **b**), plotted as $v / T$ vs $T$, with doping values as indicated. The arrows indicate the value of $T_v$ at each doping. The horizontal error bars indicate the uncertainty in determining the location of $T_v$. These $T_v$ values are listed in Table S1 and plotted with their error bars in Fig. 2. During the measurement of the $y = 6.998$ *b*-axis sample, data between 155 and 250 K was lost. As the data below 155 K was clearly sufficient to see the pseudogap-related drop in $v / T$ and define $T_v$ unambiguously, we did not repeat the measurement. In order to calculate the anisotropy difference $D(T) \equiv (v_a - v_b) / T$ up to 200 K (Fig. S9b), we interpolate the data linearly between 155 and 250 K, as shown by the red dashed line in **f**.

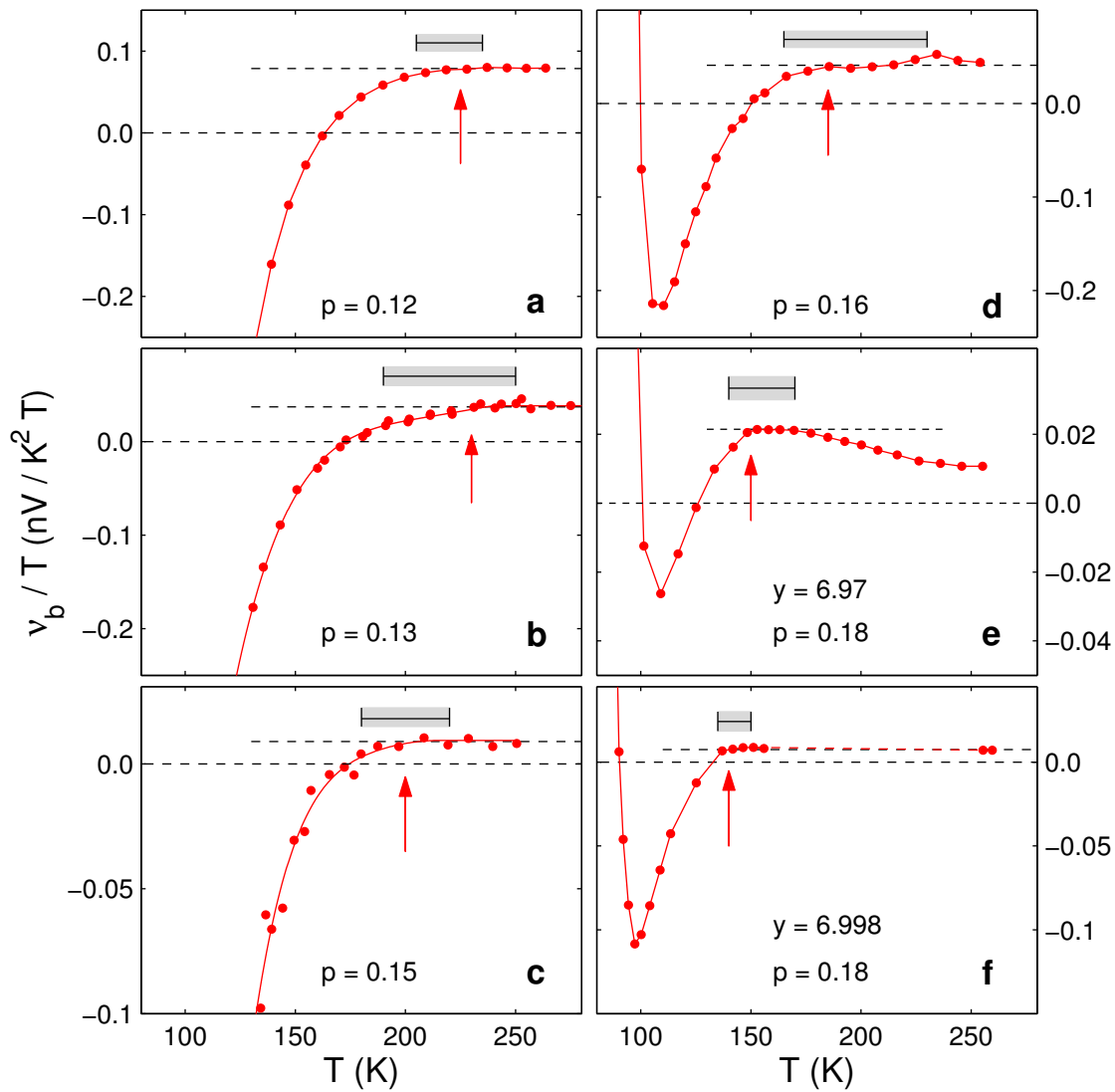

## Figure S2 | Nernst coefficient of *a*-axis samples (Δ*T* // *a*).

**a – f,** Nernst coefficient *v* of *a*-axis YBCO samples (Δ*T* // *a*) measured in a magnetic field *B* = 15 T (10 T for the *p* = 0.13 sample in **c**), plotted as *v* / *T* vs *T*,  of *a*-axis samples (Δ*T* // *a*) with doping values as indicated. The arrows indicate the value of $T_v$ at each doping. The horizontal error bars indicate the uncertainty in determining the location of $T_v$. These $T_v$ values are listed in Table S1 and are plotted with their associated error bars in Fig. 2.

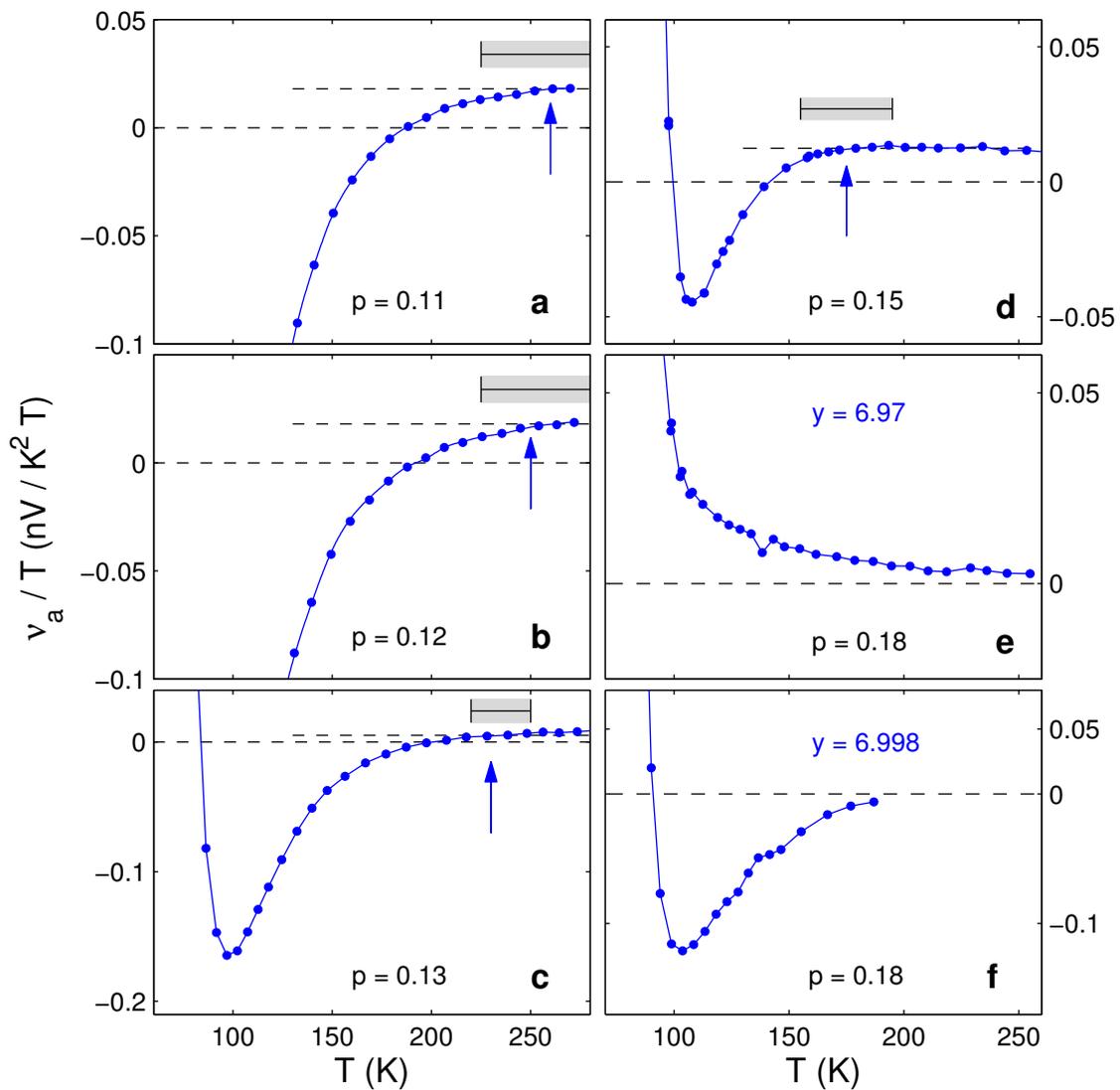

**Figure S3 | Comparison of *a*-axis and *b*-axis Nernst signals.**

**a – f,** Nernst coefficient *v* of YBCO measured in a magnetic field *B* = 15 T (10 T for the *p* = 0.13 sample in **c**), plotted as *v* / *T* vs *T*, comparing directly the *a*-axis (blue) and *b*-axis (red) signals at each doping. A pronounced anisotropy is observed at all dopings, with $v_b$ becoming much more negative than $v_a$ at low temperature, except for the 6.998 samples (in **f**), where the highly conducting chains contribute an anisotropy in the opposite direction, causing $v_a$ / *T* to be anomalously negative, even at *T* > $T_v$.

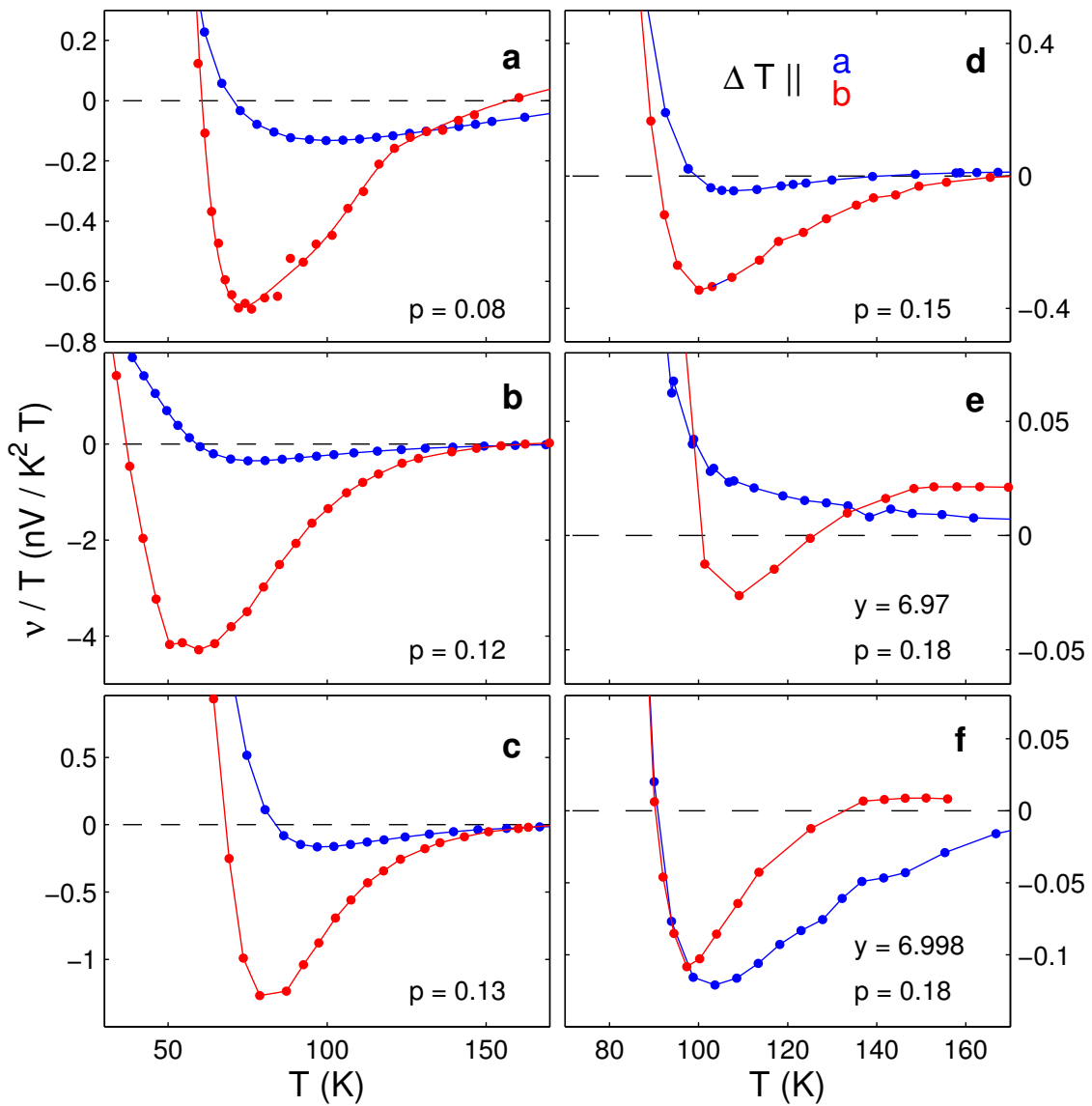

**Figure S4 | Definition of *T_ρ* and comparison of ρ and ν / *T*.**

**a,** Resistivity of YBCO *p* = 0.13 for *J // a* (from ref. 19). The line is a linear fit to the data at high temperature. **b,** Difference between the data and the fit in **a**, $\Delta\rho_a = \rho_a$ − fit. The temperature below which $\rho_a(T)$ deviates from linearity, or $\Delta\rho_a(T)$ deviates from zero, is defined as $T_\rho$. **c,** Resistivity for *J // a* in the *p* = 0.13 sample studied here. Comparison with panel **a** shows excellent agreement with the data of Ando *et al.* (ref. 19). **d,** In this panel, we compare the drop in resistivity (green) with the drop in the Nernst coefficient (blue) measured on the same sample (*a*-axis *p* = 0.13). We plot $\Delta\rho_a$ calculated from the data and fit in panel **c** and $\Delta\nu / T$, the difference between the ν / *T* data in Fig. S2c and the constant dashed-line fit at high temperature (Fig. S2c). $\Delta\nu / T$ is shown for $\Delta T // a$ (blue circles; data from Fig. S2c) and $\Delta T // b$ (red circles; data from Fig. S1b). The value of $T_\nu$ for $\Delta T // a$ and $\Delta T // b$ is shown as arrows (from Figs. S1b and S2c).

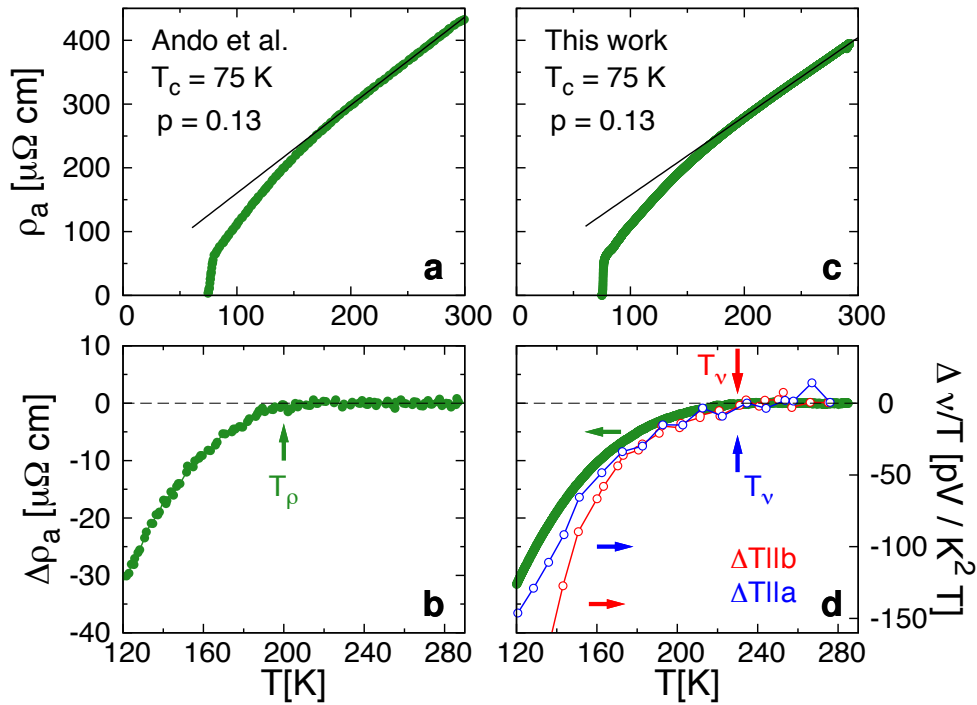

**Figure S5 | Anisotropy of the Nernst signal: difference.**

Difference in the Nernst signal of YBCO between $\Delta T \parallel a$ (data in Fig. S2) and $\Delta T \parallel b$ (data in Fig. S1) measured at a given doping, defined as $D(T) \equiv (\nu_a - \nu_b) \,/\, T$, for dopings as indicated. The inset of panel **a** is a zoom on the $p = 0.12$ data at high temperature. The arrows show the location of $T_\nu$ (from $b$-axis data in Fig. S1). Upon cooling, the increase in $D(T)$ above its very small nearly flat value at high temperature is seen to start precisely at $T_\nu$ in all cases, showing that the onset of the pseudogap phase is causing the anisotropy. The colour-coded dashed lines are linear fits to the data above $T_\nu$; the fact that they have a slight downward slope may reflect a small contribution from CuO chains, better seen in the 6.998 samples (Fig. S9b). Note that the slow initial rise in $D(T)$ below $T_\nu$ is due to the slow initial rise in the signal itself (Fig. S7).

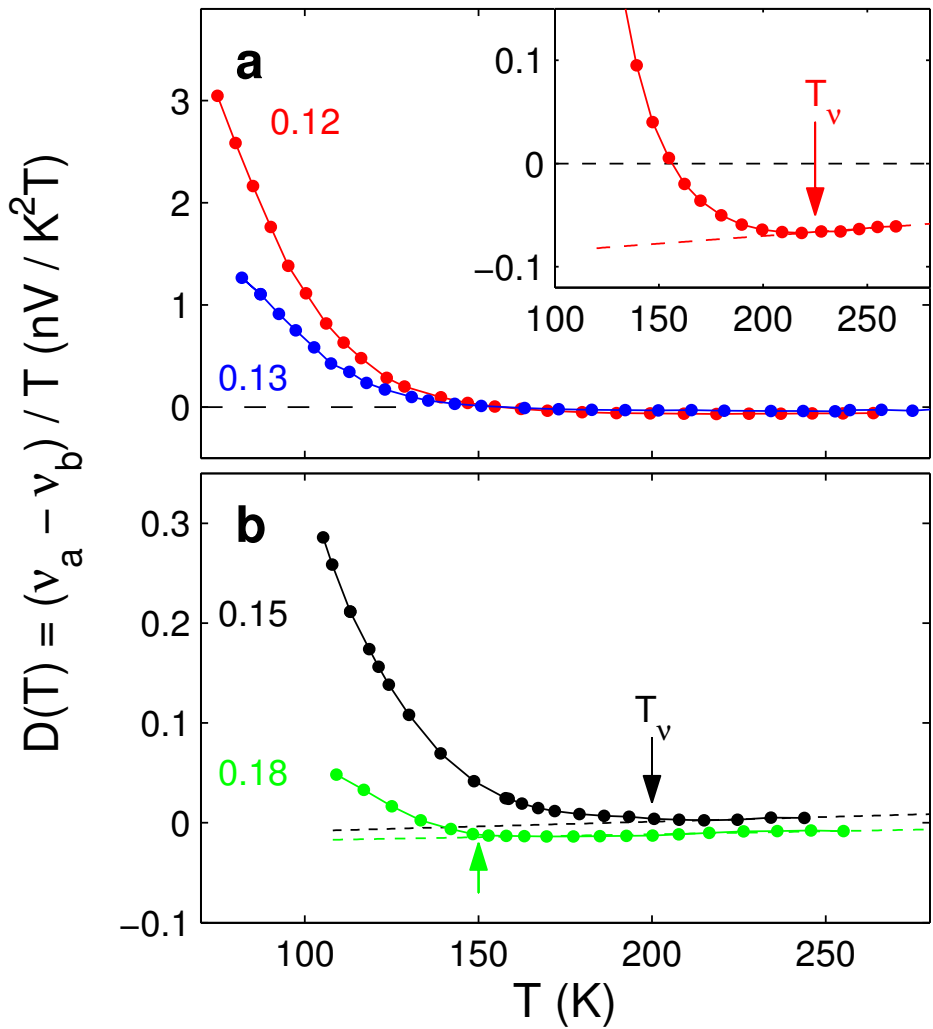

**Figure S6 | Anisotropy of the Nernst signal: ratio.**

Anisotropy of the Nernst signal compared with the corresponding anisotropy of the conductivity, both plotted as ratios: $v_b / v_a$ (dots) and $\sigma_b / \sigma_a$ (curve), respectively. The separate data for $v_a$ and $v_b$ are shown in Fig. S3. **a,** For $p = 0.08$, we see that both ratios rise with decreasing temperature, roughly tracking each other (but with $v_b / v_a$ being considerably larger). The fact that $\sigma_b / \sigma_a \to 1$ at high temperature shows that the conductivity of CuO chains is negligible at this doping, as previously demonstrated[5]. This implies that the large anisotropy in the Nernst signal is a property of the $CuO_2$ planes. **b,** At $p = 0.12$, the chains now conduct[5]. While they dominate the anisotropy in $\sigma$ and completely modify the temperature dependence of $\sigma_b / \sigma_a$ (with respect to that seen at $p = 0.08$), the behaviour of $v_b / v_a$ remains much the same as for $p = 0.08$. There is a $\pm$ 20 % error bar on $v_b / v_a$ (shown for 90 K) from the $\pm$ 10 % uncertainty on each of $v_b$ and $v_a$.

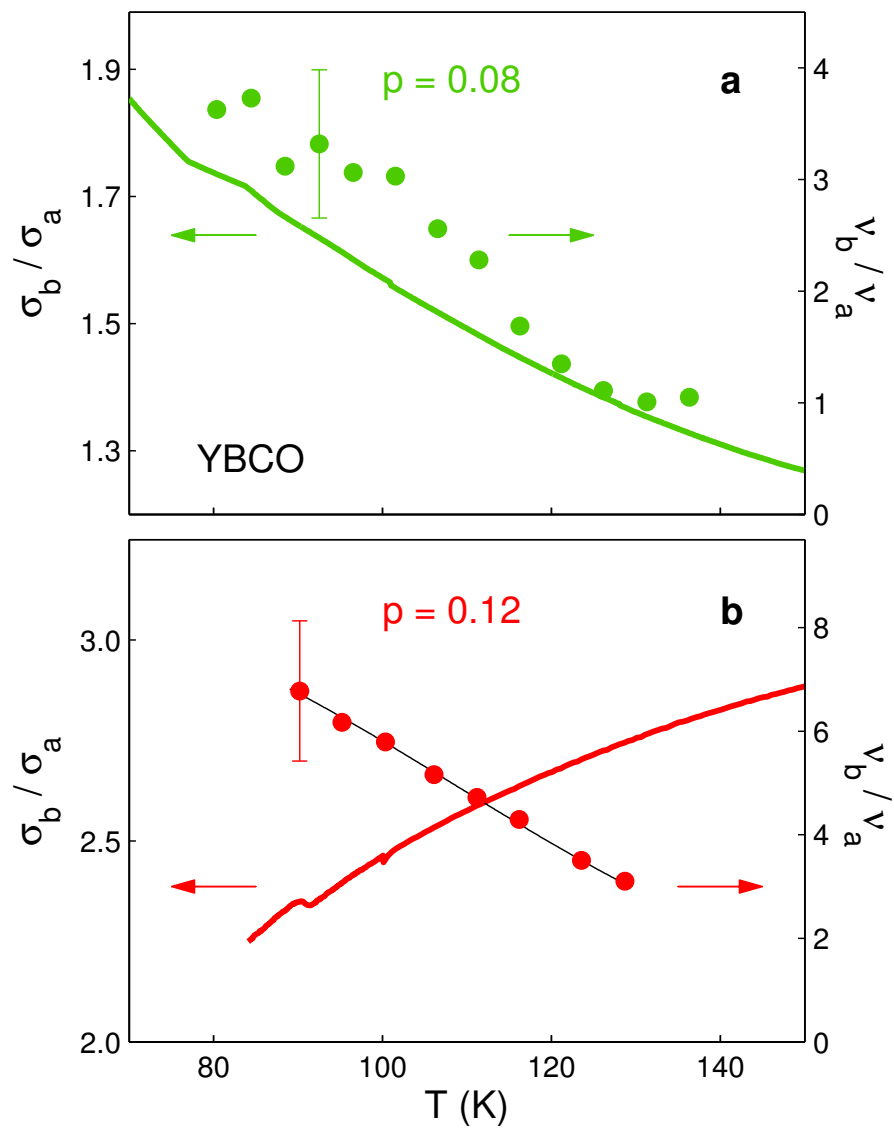

**Figure S7 | Anisotropy of the Nernst signal: difference vs sum.**

The *a-b* anisotropy of the Nernst coefficient $v$ can be displayed as a ratio, $v_b / v_a$ (as in Fig. S6) or as a difference, $D(T) = (v_a - v_b) / T$ (as in Fig. S5). Above $T_v$, $D(T)$ is very small but not quite zero, and it rises dramatically below $T_v$. In order to display purely the pseudogap-induced anisotropy, we can subtract the small background anisotropy, and plot either $D(T) - D(T_v)$, as in Fig. 4a, or more precisely $D(T) - D_0(T)$, as in panel **b**, where $D_0(T)$ is a linear fit to $D(T)$ above $T_v$ (see panel **d**). However, $D(T)$ is not a transparent measure of the anisotropy because its growth is dominated by the dramatic growth in the underlying Nernst signal $v$ itself. A more revealing quantity to look at is the ratio of difference over sum, or $D(T) / S(T) = (v_b - v_a) / (v_b + v_a)$, where $S(T) \equiv - (v_b + v_a) / T$. This quantity can be viewed as a "nematic order parameter" (ref. 34), analogous to the equivalent ratio derived from the resistance, $(R_x - R_y) / (R_x + R_y)$, used as a measure of nematicity in 2D electron gases and $Sr_3Ru_2O_7$ (ref. 34). Using the raw data for $v_a$ and $v_b$ in YBCO at $p = 0.12$ (from Fig. S3b), this ratio is plotted in Fig. 4b and panel **e** (full red dots). The degree of nematicity is large at low temperature, roughly 0.8 at 90 K, for an absolute maximum of 1.0. However, because both $(v_a - v_b)$ and $(v_a + v_b)$ change sign near 150 K (panels **c** and **d**), it becomes meaningless to plot $(v_b - v_a) / (v_b + v_a)$ above 120 K or so. We can avoid this complication by measuring $D(T)$ and $S(T)$ relative to their value at $T_v$, *i.e.* by plotting $[D(T) - D(T_v)] / [S(T) - S(T_v)]$, as in Fig. 4b and panel **e** (open circles). (For comparison, we also plot $[D(T) - D_0] / [S(T) - S_0]$ and $[D(T) - D_0] / [S(T) - 2S_0]$ in panel **e**, with $S_0$ a small constant offset; see panel **c**.) Note that the uncertainty becomes large as $T \to T_v$, where the denominator approaches zero, so the detailed rise just below $T_v$ is not known. At low temperature, however, $[D(T) - D(T_v)] / [S(T) - S(T_v)] \approx (v_b - v_a) / (v_b + v_a)$ is well-defined and accurately known.

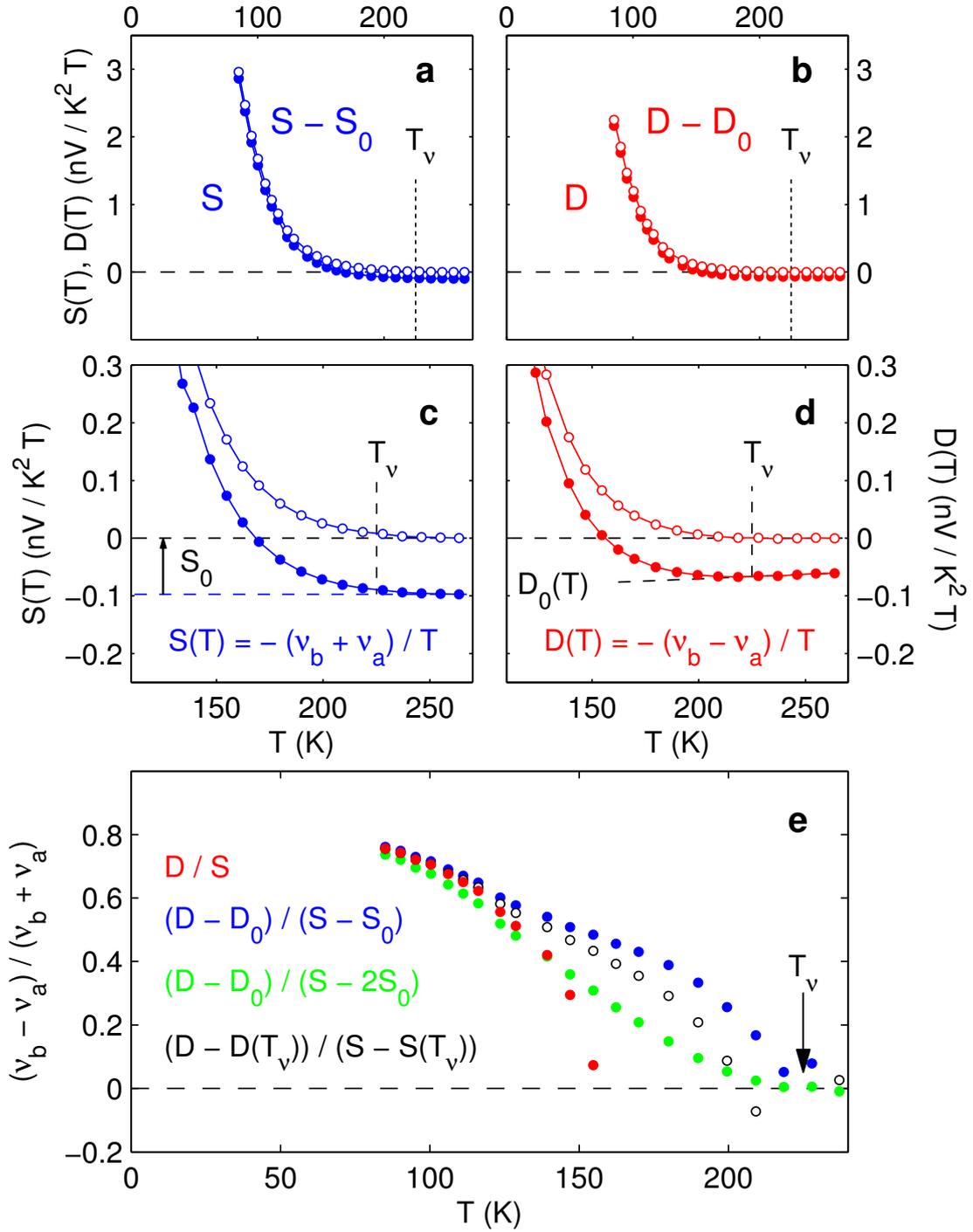

**Figure S8 | Conductivity anisotropy in samples with *y* = 6.97 vs *y* = 6.998.**

Anisotropy of the in-plane conductivity $\sigma(T)$ of YBCO at $p \approx 0.18$, for samples with oxygen content *y* = 6.97 (blue) and *y* = 6.998 (red). **a,** Anisotropy ratio $\sigma_b / \sigma_a$. A value of 4.7 reached near 150 K is the largest anisotropy ratio reported to date, indicating a high level of order and purity in the CuO chains of these 6.998 samples. **b,** Anisotropy difference $\sigma_b - \sigma_a$, a direct measure of the chain conductivity. By going from 3% oxygen vacancies in the CuO chains of the *y* = 6.97 samples to 0.2% vacancies in the *y* = 6.998 samples, the conductivity of chains is enhanced by a factor 4.

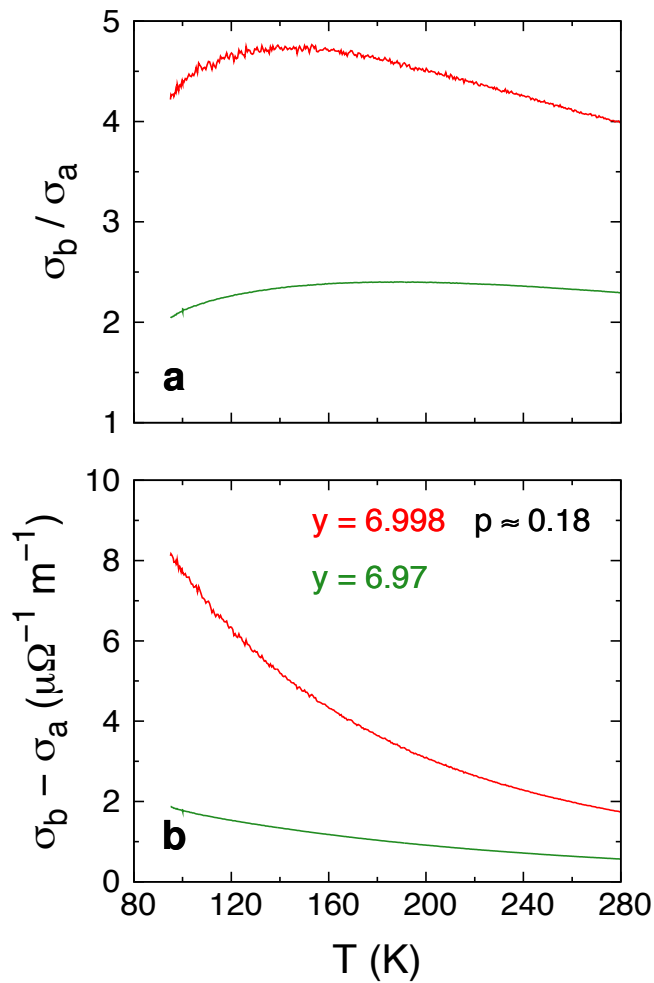

**Figure S9 | Chain contribution to the Nernst anisotropy.**

Here we compare the two pairs of samples whose conductivity anisotropy is shown in Fig. S8, with $y = 6.97$ (green) and $y = 6.998$ (brown). **a,** Chain resistivity of the 6.998 samples, defined as $\rho_{chain} \equiv 1 / (\sigma_b - \sigma_a)$, plotted vs $T^2$. $\rho_{chain}$ is seen to exhibit a perfect $T^2$ dependence from $T_c$ to 300 K, known to be characteristic of chains both in $YBa_2Cu_3O_y$ (ref. 35) and in $YBa_2Cu_4O_8$ (ref. 36). Note that the $T^2$ dependence persists unperturbed through $T_v$ (arrow), evidence that chains are unaffected by the onset of the pseudogap phase. **b,** Anisotropy difference in the Nernst signal, $D(T) \equiv v_a / T - v_b / T$, plotted as $D(T) - D(T_v)$ versus $T / T_v$. For a given pair of samples, we use the value of $T_v$ for the $b$-axis sample; the same is true for Fig. 4a. The non-zero downward-sloping background in the 6.998 data above $T_v$ is a clear manifestation of the enhanced chain conductivity. The nearly flat background above $T_v$ in the 6.97 samples, and indeed at all other dopings (see Fig. 4a and Fig. S5), shows that chains make a negligible contribution to the Nernst anisotropy above $T_v$ unless they are extremely conducting, as in the 6.998 samples. Assuming that the chain-induced background in the 6.998 extends smoothly below $T_v$, as sketched by the dashed line, we can subtract that background (dashed line) from the 6.998 data to get the chain-free data shown in panel **c**. The resulting chain-free anisotropy is then seen to be the same for both pairs of samples. Support for the assumption that the chain contribution extends smoothly through $T_v$ comes from the fact that the chain conductivity goes through $T_v$ unperturbed, as shown for $YBa_2Cu_3O_{6.998}$ in panel **a**. The same is true for $YBa_2Cu_4O_8$ through $T^*$ (ref. 36).

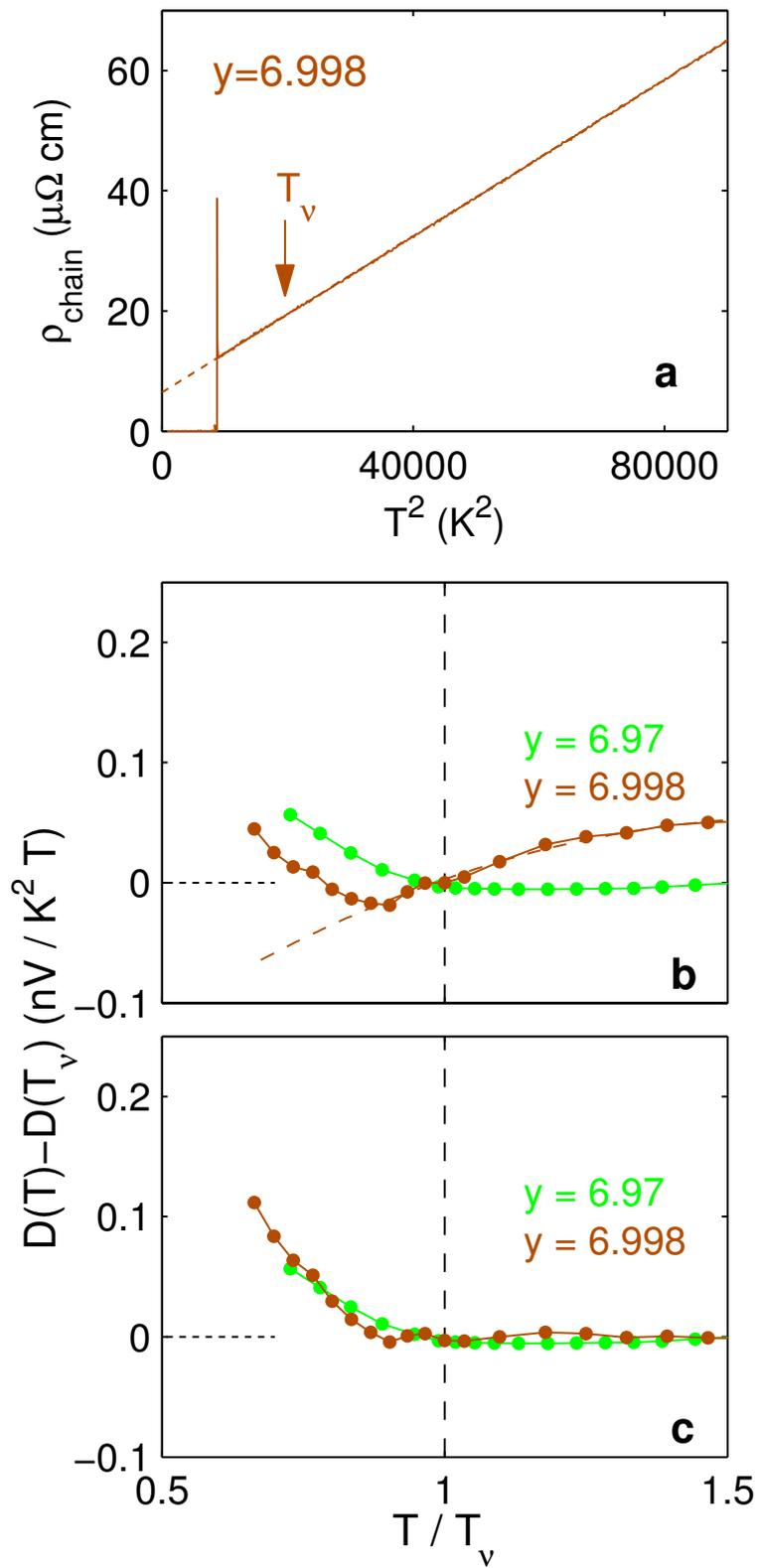